# How high-status women promote repeated collaboration among women in male-dominated contexts

Huimin Xu
School of Information, University of Texas at Austin, Austin, TX 78701, USA
Jamie Strassman
School of Business, The University of Texas at Austin, Austin, TX 78705, USA
Ying Ding
School of Information, University of Texas at Austin, Austin, TX 78701, USA
Steven Gray *
School of Business, The University of Texas at Austin, Austin, TX 78705, USA
Maytal Saar-Tsechansky *
School of Business, The University of Texas at Austin, Austin, TX 78705, USA

*Corresponding authors:  steven.gray@mccombs.utexas.edu,
Maytal.Saar-Tsechansky@mccombs.utexas.edu

## Abstract

Male-dominated contexts pose a dilemma: they increase the benefits of repeated collaboration among women, yet at the same time, make such collaborations less likely. This paper seeks to understand the conditions that foster repeated collaboration among women versus men in male-dominated settings by examining the critical role of status hierarchies. Using collaboration data on 8,232,769 computer science research teams, we found that when a woman holds the top-ranking position in a steep status hierarchy, other women on that team are more likely than men to collaborate again, as compared to when the hierarchy is flat, and compared to when men occupy the top-ranking position. In steep hierarchies, top-ranking women but not top-ranking men foster conditions in which junior women are more likely to collaborate again than junior men of similar status levels. Our research suggests that whereas status hierarchies are especially detrimental to repeated collaboration among underrepresented individuals, top-ranking women in steep status hierarchies mitigate these negative impacts between women in male-dominated settings.

## Significance Statement

Using scientific collaboration data from 8,232,769 computer science research teams, we found that when a team has a steep status hierarchy and a woman holds the top position in that hierarchy, other women on that team are more likely to collaborate again as compared to men. However, the opposite is true in the presence of flat hierarchies or when men hold the top position. Even though status hierarchies typically trigger negative relational dynamics among underrepresented individuals, we found that top-ranking women can override these negative dynamics and foster productive relationships among women in male-dominated contexts.

## Introduction

*The glass ceiling will go away when women help other women break through that ceiling*, Indra Nooyi

*Both Emmanuelle and I felt proud of our gender that morning and just happy that we were sending a message collectively to girls and others who have felt excluded from the STEM fields that their work can be recognized,* Jennifer Doudna on winning the Nobel Prize in Chemistry with collaborator Emmanuelle Charpentier

Research suggests that repeated collaboration between women, particularly in male-dominated settings, drives a host of societal, organizational, and individual-level benefits, such as increased innovation (1), firm performance (2), and female advancement (3). In settings where men make up most of the labor pool, women are often conferred comparatively less status (4) and as a result, are relegated to the outskirts of both formal and informal networks (5–7). To gain access to the task-related information (8), financial resources (9, 10), gender-specific advice (11), visibility (12), and psychosocial support (2) necessary to excel in their day-to-day activities, research suggests that women benefit from repeated collaborations with *one another*. Men, on the other hand, who typically enjoy high status and centrality in organizational networks within male-dominated settings (6), need not rely on same-gender others to successfully carry out their work (13, 14). In short, in male-dominated contexts, it is uniquely critical that women, even more so than men, rely on repeated same-gender connections (13, 14). However, it is in these very settings where woman-to-woman relationships are most fraught (15) and man-to-man relationships are more likely (16). In male-dominated contexts, women are often hesitant to regularly collaborate with each other because it can incur greater levels of status loss (17). By frequently affiliating with other women, they risk drawing undue attention to their gender, making it a more likely basis for judgment (18). Additionally, in such settings, high-status roles are seen as scarce for women, increasing intra-gender competition among women, in particular (19, 20). As such, professional women face a *collaboration dilemma*: in the spaces where they could benefit most from regular collaboration, their ability to do so is uniquely limited. That is, in the contexts in which women are most underrepresented, their need to frequently collaborate is heightened, yet men, more so than women, likely engage more frequently in such relationships.

To enable women's advancement, we must better understand the conditions that nurture repeated collaboration among women (as compared to men) within the male-dominated settings that have historically precluded it. Given the central role that status—formally defined as the degree to which an individual is conferred respect, admiration, and influence from others (21)—plays in this collaboration dilemma (18), we investigate how status hierarchies affect repeated woman-to-woman (versus man-to-man) collaboration in a male-dominated context. We do so by analyzing 8,232,769 research teams within a particular male-dominated context—computer science (CS). Collaboration among women in STEM research has been shown to be particularly beneficial (1, 12) yet unlikely (16), making CS an ideal research context. Within this context, we examine how the steepness of a team's status hierarchy, the gender of the top-ranking team members, and the dyadic differences in status among other team members predict same-gender author re-collaboration on future research projects.

**Results**

# High-status women in hierarchical teams foster woman-to-woman repeated collaboration

Drawing on data from 8,232,769 gender-diverse teams in the CS field, we examine how the steepness of a team's status hierarchy, the gender of the top-ranking scientists, and the dyadic difference among other team members influence team members' likelihood of repeated collaboration. Specifically, we look at how these factors influence *same-gender* re-collaboration. Our data is derived from the bibliographic details of papers from prominent CS journals and conferences, and we consider those papers published by a *team* if they had at least two authors. As illustrated in Fig 1, re-collaboration captures whether a same-gender pair within the team chose to work together again in the future. Team status hierarchy is quantified as the Gini coefficient of all team members' H-indexes before publishing the focal paper (22) The H-index is a metric used to measure both the productivity and citation impact of a scientist's publications. The Gini coefficient is a measure of inequality, often used to assess wealth or income distribution, but here it is used to quantify the distribution of professional status within each team (23–25). We examine the impact of the gender of the top-ranking scientists in two ways: 1) we record whether a woman or a man held the highest status within the team (Fig 2) and 2) whether women versus men collectively held higher status relative to the team (Fig 3).

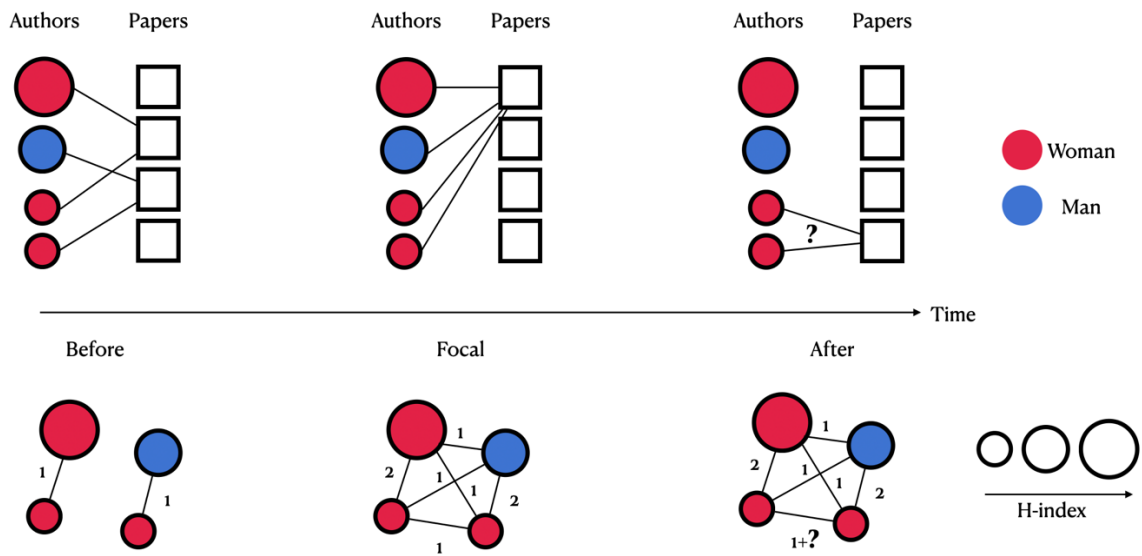

**Fig 1.** A case of repeated collaboration between scientist pairs. Before, a higher H-index woman collaborated with a lower H-index woman once, and a higher H-index man collaborated with a lower H-index man once. In the focal paper, these four researchers form a new team. Our research asks how the team status hierarchy and the status of women in that hierarchy influence scientists of the same gender and varying status levels (e.g., two lower H-index women) to collaborate again on a future project.

Our analyses regarding the impact of team status hierarchy and whether a woman versus a man held the top-ranking position were performed using a logistic regression model with the formula: *Re-collaboration ~ Team Hierarchy * Gender Pairs * Gender with the Highest Status + Expertise Similarity + Prior Collaboration Frequency + Women Number + Mean Team H-index + Year + Team Size (formula 1)*. In the regression model, the basic unit is the same-gender pairs within teams, including woman-to-woman and man-to-man pairs. The dependent variable is their predicted re-collaboration probability. While controlling for expertise similarity between scientist pairs, prior collaboration frequency between scientist pairs, the number of women in the team, mean team H-index, the publication year of the focal paper, and team size we find that the three-way interactive effect of *Team Hierarchy * Gender Pairs * Gender with Highest Status* is significant ($p < 0.001$, Table 2, Model 1, SI and Appendix). As seen in Fig 2a, when a woman holds the highest status position within a *flat* team (low Gini coefficient), men are more likely than women to collaborate again. Critically, however, when a woman has the highest status within a *hierarchical* team (high Gini coefficient), women team members demonstrate a higher probability of repeated collaboration than men. Yet, as seen in Fig 2b, when a man has the highest H-index, men are always more likely to repeatedly collaborate, regardless of team hierarchy. Finally, it is worth noting that as demonstrated in Fig 2c and Model 1 (SI and Appendix), when a woman has the highest status within a *hierarchical* team, women team members are not only more likely to re-collaborate as compared to men, but they are also more likely to re-collaborate as compared to women working in hierarchical teams with high-ranking men.

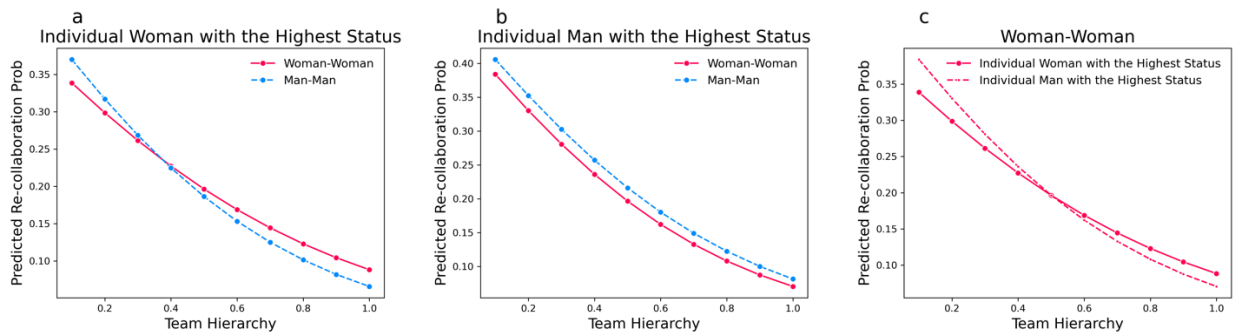

**Fig 2.** Comparing how a woman/man with the highest status influences repeated collaboration **(a)** Woman-woman pairs are more likely to re-collaborate than man-man pairs when a woman has the highest H-index in hierarchical teams. **(b)** Man-man pairs are more likely to re-collaborate than woman-woman pairs when a man has the highest H-index, regardless of team hierarchy. **(c)** Woman-woman pairs are more likely to re-collaborate in hierarchical teams when a woman has the highest H-index as compared to when a man has the highest H-index. Team hierarchy is drawn for every 0.1 increase, ranging from 0.1 to 1. Gender Pairs are marked with two different colors, woman-woman (red), and man-man (blue). The model results are shown in Table 2 Model 1 (SI and Appendix).

Our analyses regarding the impact of team status hierarchy and whether women versus men collectively held higher team status were again conducted using a logistic regression model with the formula: *Re-collaboration ~ Team Hierarchy * Gender Pairs * Collective Women Status + Expertise Similarity + Prior Collaboration Frequency + Women Number + Mean Team H-index + Year + Team Size (formula 2).* In this model, the only different independent variable from formula 1 is Collective Women Status, the average H-index for women collectively relative to the average team H-index. Collective Women with Higher Status (Fig 3a) and Collective Women with Lower Status (Fig 3b) separately represent the Mean + SD and Mean - SD of the continuous variable Collective Women Status. While controlling for all the confounding variables, we found that the three-way interactive effect of *Team Hierarchy * Gender Pairs * Collective women Status* is significant ($p < 0.001$ In Model 2, Table 2, SI and Appendix). As seen in Fig 3a, when women collectively have higher status in hierarchical teams, woman-to-woman pairs exhibit a significantly higher propensity for re-collaboration compared to men. That is, when the average H-index for women is higher than the average team H-index and the team hierarchy is steep, women pairs have a higher probability of re-collaborating compared with men pairs. However, this effect disappears when the average H-index for women is lower than the average team H-index (Fig 3b). When women, on average, have lower status, men consistently have a higher probability of re-collaborating compared to women. Here again, it is important to note that women working within hierarchical teams where women collectively hold higher status are more likely than women working within hierarchical teams where women collectively hold lower status to re-collaborate (Fig 3c).

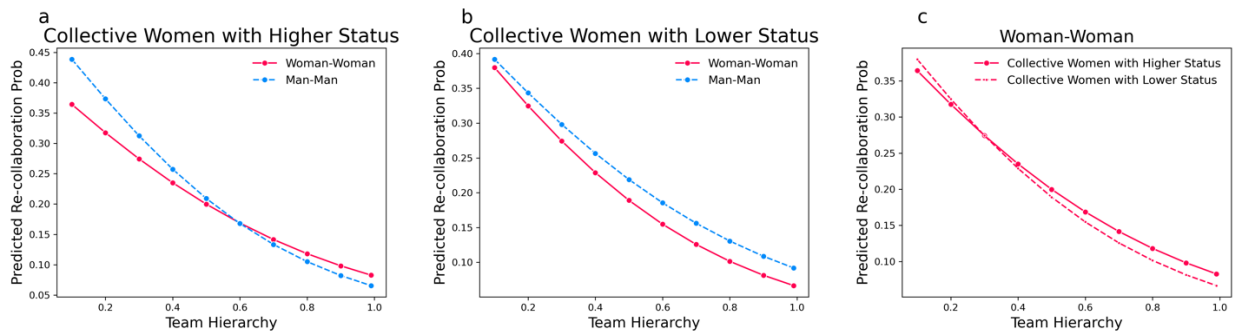

**Fig 3.** Comparing how the average status of women versus that of men influences repeated collaboration **(a)** Women pairs are more likely to re-collaborate than men pairs when the average H-index for women is higher than the average team H-index in hierarchical teams. **(b)** Men pairs are more likely to re-collaborate than women pairs when the average H-index for women is lower than the average team H-index, regardless of team hierarchy. **(c)** Woman-woman pairs are more likely to re-collaborate in hierarchical teams when women collectively have a higher H-index as compared to when women collectively have a lower H-index. Team hierarchy is drawn for every 0.1 increase, ranging from 0.1 to 1.

Gender Pairs are marked with two different colors, woman-woman (red), and man-man (blue). The model results are shown in Table 2 Model 2 (SI and Appendix).

Other interesting findings include that hierarchical structures are less likely to contribute to scientists' repeated collaboration than flat structures ($p < 0.001$); large teams are less likely to maintain relationships than small teams ($p < 0.001$); prior collaboration experience can increase the re-collaboration probability between scientists ($p < 0.001$); when teams have higher H-indices on average, scientists are more likely to re-collaborate ($p < 0.001$). See Table 2 Model 1 and 2 (SI and Appendix).

**High-status women in hierarchical teams foster repeated collaboration among junior women of similar status**

We also investigate how the dyadic status differences between scientists contribute to the interactive effects found above. Specifically, we ran a three-way interactive effects model where *formula = Re-collaboration ~ Team Hierarchy * Gender pairs * Dyadic Status Difference + Expertise Similarity + Prior Collaboration Frequency + Women Number + Mean Team H-index + Year + Team Size (formula 3).* This formula is applied to two different situations: 1) when a woman has the highest status (Fig 4a), and 2) when a man has the highest status (Fig 4b). Dyadic Status Difference is defined as the H-index difference between scientist pairs, divided by the mean value 3, with high (>= mean) and low (< mean) categories. As seen in Fig 4a, in teams where a woman has the highest status, women pairs with high status differences are always more likely to re-collaborate than men with high status differences, regardless of team hierarchy. But when status differences are lower for scientist pairs, an interactive effect emerges (Table 3 Model 1, B = 0.48, SE = 0.033, $p < 0.001$, SI and Appendix). When dyadic status differences are low, women have higher repeated collaboration rates than men in hierarchical (but not flat) teams. As seen in Fig 4b, in teams where a man has the highest status, men with low status differences are always more likely than women with low status differences to re-collaborate, regardless of team hierarchical structure. Interestingly, pairs with low dyadic status differences, on average, are of lower status and more junior. This indicates that the presence of high-status women in hierarchical teams can facilitate repeated collaboration among junior women with lower H-indices, thereby contributing to their development via future collaboration opportunities.

The impact of women's collective status on whether women and men of varying status differences re-collaborate exactly mirrors those displayed in Fig 4a-b (see Fig 4c-d). The interactive effect in Fig 4c is significant (Table 3 Model 3, B = 0.53, SE = 0.026, $p < 0.001$, SI and Appendix).

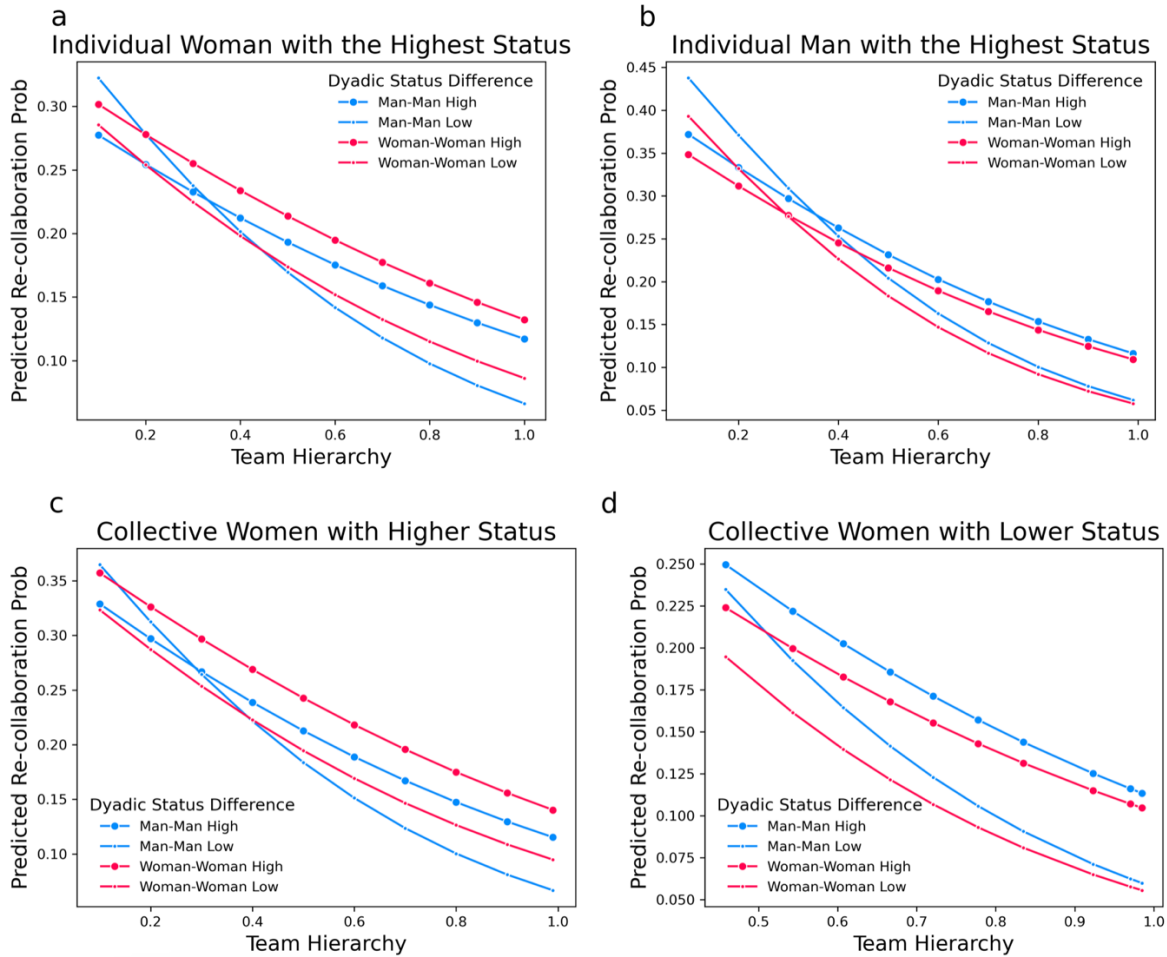

**Fig 4.** Comparing how dyadic status difference between researchers influences repeated collaboration when women have higher or lower status. **(a, c)** In situations of low dyadic status difference, women pairs are more likely to re-collaborate than men pairs when a woman has the highest status/woman collectively have higher status in hierarchical (but not flat) teams. In situations of high dyadic status difference, woman pairs are more likely to re-collaborate than men pairs when a woman has the highest status/women collectively have higher status, regardless of hierarchical structure. **(b, d)** In situations of low dyadic status difference, women pairs are less likely to re-collaborate than men pairs when a man has the highest status/men collectively have higher status, regardless of hierarchical structure. In situations of high dyadic status difference, woman pairs are again less likely to re-collaborate than men pairs when a man has the highest status/men collectively have higher status, regardless of hierarchical structure. Team hierarchy is drawn for every 0.1 increase, ranging from 0.1 to 1. Gender Pairs are marked with two different colors, woman-woman (red), and man-man (blue). Dyadic power differences are marked with two different shapes, high (dots) and low (lines). The model results are shown in Table 3 (SI and Appendix).

## Discussion

In this paper, we identified a collaboration dilemma present in male-dominated contexts and aimed to pinpoint the conditions where it broke down. Using data from 8,232,769 gender-diverse teams in the CS field, we found that on average, women are less likely than men to repeatedly collaborate. However, one condition flips this relationship and uniquely fosters woman-to-woman (as compared to man-to-man) re-collaboration: when a team has a steep status hierarchy and a woman or women, sit at the top of that hierarchy. More specifically, we found that top-ranking women in steep status hierarchies are especially adept at promoting repeated collaboration among junior women team members who have *similar* status levels. Whereas prior research has suggested that team members from underrepresented groups (e.g., women in male-dominated professions) with similar status levels should be *less* likely to have positive relationships than majority group members (19, 26, 27), we instead find the opposite, so long as the team is led by a woman and has a steep status hierarchy.

Collectively, our results offer a resolution to the collaboration dilemma and suggest that women leaders in male-dominated contexts can help safeguard the highest risk relationships: women-to-women relationships in steep status hierarchies, particularly those between women who have similar status levels. Our findings point to the notion that although steep status hierarchies are typically associated with greater intra-team competition (28), when women hold the top positions in such hierarchies, it may afford them the unequivocal respect and influence necessary for them to effectively, and positively, impact the collaborative climate within the team, particularly as it relates to fostering the development of other woman-to-woman relationships. Future research should explore this potential intervening mechanism.

## Materials and Methods

**Dataset**: We choose the CS field from the OpenAlex dataset (https://docs.openalex.org/), which records bibliographic details of papers from prominent CS journals and conferences. Until 2023 May, the dataset encompasses 57,231,187 CS papers. For this study, we define teams as those with a minimum of two authors in a publication, and a total of 30,440,575 teams. Since we investigate how the status of women within teams influences scientists' repeated collaboration, we keep 8,232,769 (27%) gender diverse teams with at least one woman.

**Gender Prediction:** To explore the relationship between team hierarchy and gender diversity, we need to identify the information of team members' gender. To achieve this, we utilized a Python package developed by Liang & Acuna (2021) that employs deep learning transformers to predict gender based on authors' first name. Gender is classified into three categories: men, women, and unknown with specific probability. We label scientists with the largest probability of gender category. After excluding authors with unknown gender, there are 11,416,286 identified male scientists and 4,462,318 identified female scientists. The identified female and male scientist ratio is 0.39:1 According to National Center for Educational Statistics, around 20-40% women are awarded the bachelor's degree in the computer science field during the past 50 years (https://ischoolonline.berkeley.edu/blog/women-computing-computer-science/). It might help us validate the results of identifying women when we assume researchers publishing papers at least receive the bachelor's degree.

**Dependent Variables:** Predicting repeated collaboration, whether the same-gender scientists will repeat collaboration after publishing one paper, 1 is yes, 0 is no. 10,863,704 (18%) pairs choose to continue collaborating. Since we are intended to study whether scientists will repeat collaborate after publishing one paper, scientist pairs in teams are the basic unit of our observation.

**Independent Variables:**
Team Hierarchy: We calculate the Gini coefficient of H-index within the team members before publishing the focal paper. H-index is a metric considering both the productivity and impact, which can represent the scientific status of a scientist. $H=\max \{h : c_h \geq h\}$, Where $H$ is the H-index of a scientist, $c_h$ is the number of citations of the $h^{th}$ paper in the CS field, when papers are ranked by citation count. If a team member does not have any publications in the CS field before, his H-index is 0. When all the researchers in a team are new in the CS field and do not have any publications before, the team hierarchy is not considered in our paper.
Gender Pairs: There are 6,533,063 woman-woman, 28,280,260 man-man, 26,469,739 woman-man pairs in teams with at least one woman. In this paper, we only focus on woman-woman and man-man pairs.
Collective Women Status: We measure it with average women H-index - average team H-index, meaning the average women status relative to the average all team members' status.
Gender with the Highest Status: Whether woman or man is the individual with the highest H-index in teams. After excluding those teams where the individual with the highest status is gender unknown, there are 1,441,624 teams where the individual with the highest status is woman, and 3,068,857 teams where the individual with the highest status is man.
Dyadic Status Difference: The H-index difference between two scientists in pairs.
The descriptive information and correlations of these variables are shown in Table 1 (SI and Appendix).

**Confounding Variables:**
Expertise similarity: The expertise similarity between scientist pairs is qualified by keywords in previous papers within 10 years. It is the ratio between the overlap keyword sets number and all the keyword sets number for scientist pairs before publishing the focal paper. For any scientist pair in each paper, we measure the percentage of the

keyword overlap $\frac{set(a) \cap set(b)}{set(a) \cup set(b)}$, where a and b are the keywords of two scientists before publishing the focal one within 10 years.

Prior Collaboration Frequency: the frequency before collaborating the focal paper for scientist pairs.

Women Number: the number of women in teams.

Mean Team H-index: the mean value of H-index of all the members in a team.

Year: the publication year of a focal paper.

Team Size: the number of team members.

The descriptive information and correlations of these variables are shown in Table 1 (SI and Appendix).

# SI and Appendix

Table 1. Descriptive information of variables for scientist pairs

|  | Mean | Standard deviation | 25% | 50% | 75% | 1 | 2 | 3 | 4 | 5 | 6 | 7 | 8 |
|---|---|---|---|---|---|---|---|---|---|---|---|---|---|
| 1 Team Hierarchy | 0.69 | 0.19 | 0.55 | 0.70 | 0.84 | | | | | | | | |
| 2 Collective Women Status | -0.69 | 2.21 | -1.30 | -0.33 | -0.03 | 0.13 | | | | | | | |
| 3 Dyadic Status Difference | 2.70 | 4.70 | 0.00 | 1.00 | 3.00 | -0.26 | -0.27 | | | | | | |
| 4 Expertise Similarity | 0.40 | 0.33 | 0.15 | 0.27 | 0.59 | 0.48 | 0.14 | -0.4 | | | | | |
| 5 Women Number | 114.25 | 281.38 | 1.00 | 3.00 | 9.00 | 0.56 | 0.12 | -0.21 | 0.52 | | | | |
| 6 Prior Collaboration Frequency | 0.86 | 5.01 | 0.00 | 0.00 | 0.00 | -0.2 | -0.09 | 0.07 | 0.02 | -0.07 | | | |
| 7 Mean Team H-index | 2.10 | 2.42 | 0.40 | 1.33 | 2.92 | -0.57 | -0.37 | 0.56 | -0.36 | -0.32 | 0.17 | | |
| 8 Year | 2015 | 8.41 | 2012 | 2017 | 2021 | 0.12 | -0.04 | 0.07 | 0.16 | 0.33 | 0.01 | 0.12 | |
| 9 Team Size | 588 | 1454 | 7 | 13 | 47 | 0.56 | 0.12 | -0.21 | 0.52 | 1 | -0.07 | -0.32 | 0.32 |

Table 2. Logistic regression model for predicting repeated collaboration between scientist pairs for all teams

| | Model 1: All Teams | | Model 2: All Teams |
|---|---|---|---|
| Gender Pairs [man-man] | 0.19 (0.008) *** | Gender Pairs [man - man] | 0.25 (0.005) *** |
| Gender with the Highest Status [man] | 0.24 (0.008) *** | Collective Women Status | -0.02 (0.003) *** |
| Gender Pairs [man - man] * Gender with the Highest Status [man] | -0.10 (0.011) *** | Gender Pairs [man - man] * Collective Women Status | -2.26 (0.004) *** |
| Team Hierarchy | -1.85 (0.009) *** | Team Hierarchy | -2.19 (0.007) ** |
| Team Hierarchy * Gender Pairs [man - man] | -0.50 (0.013) *** | Team Hierarchy * Gender Pairs [man - man] | -0.30 (0.007) *** |
| Team Hierarchy * Gender with the Highest Status [man] | -0.49 (0.014) *** | Team Hierarchy * Collective Women Status | 0.08 (0.006) *** |
| Team Hierarchy * Gender with the Highest Status [men] * Gender Pairs [man - man] | 0.58 (0.017) *** | Team Hierarchy * Collective Women Status * Gender Pairs [man - man] | -0.22 (0.006) *** |
| Expertise Similarity | -0.27 (0.002) *** | Expertise Similarity | -0.27 (0.002) *** |
| Women Number | -0.02 (0.000) *** | Women Number | -0.02 (0.000) *** |
| Prior Collaboration Frequency | 0.16 (0.000) *** | Prior Collaboration Frequency | 0.16 (0.000) *** |
| Mean Team H-index | 0.02 (0.000) *** | Mean Team H-index | 0.01 (0.000) *** |
| Year | -0.03 (0.000) *** | Year | -0.01 (0.000) *** |
| Team Size | -0.01 (0.000) *** | Team Size | -0.03 (0.000) *** |
| Intercept | 59.35 (0.125) *** | Intercept | 58.52 (0.117) *** |
| Observations | 30,622,876 | Observations | 34,813,323 |

Table 3. Logistic regression model for predicting repeated collaboration between scientist pairs at different levels of dyadic status difference when women have higher/lower status

| | Model 3: Individual Woman with the Highest Status | Model 4: Individual Man with the Highest Status | | Model 5: Collective Woman with Higher Status | Model 6: Collective Woman with Lower Status |
|---|---|---|---|---|---|
| Gender Pairs [man - man] | 0.23 (0.010) *** | 0.20 (0.008) *** | Gender Pairs [man - man] | 0.25 (0.008) *** | 0.38 (0.008) *** |
| Team Hierarchy | -1.61 (0.011) *** | -2.65 (0.012) *** | Team Hierarchy | -1.71 (0.011) *** | -2.69 (0.011) *** |
| Team Hierarchy * Gender Pairs [man - man] | -0.51 (0.016) *** | -0.12 (0.013) *** | Team Hierarchy * Gender Pairs [man - man] | -0.64 (0.013) *** | -0.30 (0.012) *** |
| Dyadic Status Difference | 0.03 (0.012) ** | -0.29 (0.016) *** | Dyadic Status Difference | 0.12 (0.013) *** | -0.27 (0.014) *** |
| Dyadic Status Difference * Gender Pairs [man - man] | -0.34 (0.019) *** | -0.09 (0.017) *** | Dyadic Status Difference * Gender Pairs [man - man] | -0.36 (0.015) *** | -0.19 (0.015) *** |
| Gender Pairs [man - man] * Team Hierarchy | 0.45 (0.021) *** | 1.00 (0.028) *** | Gender Pairs [man - man] * Team Hierarchy | 0.33 (0.021) *** | 0.97 (0.024) *** |
| Dyadic Status Difference * Team Hierarchy * Gender Pairs [man - man] | 0.49 (0.033) *** | 0.09 (0.029) ** | Dyadic Status Difference * Team Hierarchy * Gender Pairs [man - man] | 0.52 (0.026) *** | 0.20 (0.025) *** |
| Expertise Similarity | -0.29 (0.006) *** | 0.02 (0.003) *** | Expertise Similarity | -0.17 (0.005) *** | 0.01 (0.003) *** |
| Women Number | -0.04 (0.001) *** | -0.02 (0.000) *** | Women Number | -0.06 (0.000) *** | -0.01 (0.000) *** |
| Prior Collaboration Frequency | 0.21 (0.001) *** | 0.15 (0.000) *** | Prior Collaboration Frequency | 0.20 (0.000) *** | 0.14 (0.000) *** |
| Mean Team H-index | 0.03 (0.001) *** | 0.01 (0.000) *** | Mean Team H-index | 0.02 (0.000) *** | 0.01 (0.000) *** |
| Year | -0.04 (0.000) *** | -0.03 (0.000) *** | Year | -0.04 (0.000) *** | -0.03 (0.000) *** |
| Team Size | -0.01 (0.000) *** | -0.01 (0.000) *** | Team Size | -0.00 (0.000) *** | -0.01 (0.000) *** |
| Intercept | 83.42 (0.332) *** | 60.13 (0.137) *** | Intercept | 75.81 (0.000) *** | 57.40 (0.135) *** |
| Observations | 4,733,449 | 25,889,427 | Observations | 7,849,180 | 26,964,143 |